%% file: main.tex
 \documentclass[manuscript]{acmart}
\AtBeginDocument{%
  }

\setcopyright{acmlicensed}
\copyrightyear{2018}
\acmYear{2018}
\acmDOI{XXXXXXX.XXXXXXX}
\acmConference[Conference acronym 'XX]{Make sure to enter the correct
  conference title from your rights confirmation email}{June 03--05,
  2018}{Woodstock, NY}
\acmISBN{978-1-4503-XXXX-X/2018/06}

\usepackage{array}

\begin{document}

\title{AI Academy: Building Generative AI Literacy in Higher Ed Instructors}


\author{Si Chen}
\affiliation{%
  \institution{Notre Dame Learning \& Lucy Family Institute for Data \& Society, University of Notre Dame}
   \state{Indiana}
  \country{United States}}
\email{schen34@nd.edu}

\author{Xiuxiu Tang}
\affiliation{%
  \institution{Notre Dame Learning \& Department of Psychology, University of Notre Dame}
     \state{Indiana}
  \country{United States}}
\email{xtang8@nd.edu}

\author{Alison Cheng}
\affiliation{%
  \institution{Department of Psychology, University of Notre Dame}
     \state{Indiana}
  \country{United States}}
\email{ycheng4@nd.edu}

\author{Nitesh Chawla}
\affiliation{%
  \institution{Computer Science and Engineering \& Lucy Family Institute for Data \& Society, University of Notre Dame}
     \state{Indiana}
  \country{United States}}
\email{nchawla@nd.edu}

\author{G. Alex Ambrose}
\affiliation{%
  \institution{Notre Dame Learning, University of Notre Dame}
     \state{Indiana}
  \country{United States}}
\email{gambrose@nd.edu}

\author{Ronald Metoyer}
\affiliation{%
  \institution{Computer Science and Engineering \& Notre Dame Learning, University of Notre Dame}
   \state{Indiana}
  \country{United States}}
\email{rmetoyer@nd.edu}


\begin{abstract}
Generative AI is reshaping higher education, yet research has focused largely on students, while instructors remain understudied despite their central role in mediating adoption and modeling responsible use. We present the \textit{AI Academy}, a faculty development program that combined AI exploration with pedagogical reflection and peer learning. Rather than a course evaluated for outcomes, the Academy provided a setting to study how instructors build AI literacies in relation to tools, policies, peer practices, and institutional supports. We studied 25 instructors through pre/post surveys, learning logs, and facilitator interviews. Findings show AI literacy gains alongside new insights.
We position instructors as designers of responsible AI practices and contribute a replicable program model, a co-constructed survey instrument, and design insights for professional development that adapts to evolving tools and fosters ethical discussion.\end{abstract}



\maketitle

\input{1-intro}

\input{2-relatedworks}

\input{3-Method}

\input{4-Findings}

\input{5-Discussion}

\input{6-Conclusion}

\bibliographystyle{ACM-Reference-Format}
\bibliography{sample-base}

\input{7-appendix}

\end{document}

%% file: 1-intro.tex
\section{Introduction}

Generative AI is rapidly reshaping higher education, raising urgent questions not only about how students use tools such as ChatGPT and Gemini, but also about how instructors themselves adapt to them. While existing research emphasizes student practices in writing, coding, or problem solving \cite{kasneci2023chatgpt,zheng2023impact}, instructors remain understudied despite their central role in mediating adoption, setting classroom norms, and modeling responsible use \cite{henderson2017instructor,holstein2019student}. This gap matters because instructors’ approaches will determine not only how AI is integrated into curricula but also how future professionals learn to engage responsibly with these systems, aligning with emerging scholarship that situates AI adoption within the broader transformation of academic professions \cite{paris2025artificial}.

Commercial platforms have begun releasing educator resources \cite{microsoft2023copilot,openai2023classroom}, but these are typically product-oriented rather than pedagogy-driven. They emphasize learning how to operate tools rather than grappling with the broader challenges instructors face in balancing ethics, assessment, workload, and credibility. We take a different perspective by foregrounding instructors not only as learners within sociotechnical systems but also as designers of responsible AI practice. We ask what it means to teach the teachers AI: to design professional development that goes beyond technical tutorials and instead builds literacy while also surfacing tensions around fairness, institutional policy, and professional identity.

To explore this, we designed and deployed the \textit{AI Academy}, a faculty development program that integrated technology exploration with pedagogical reflection and peer learning. We studied its first cohort of 25 instructors through pre and post surveys, participant learning logs, and facilitator interviews. This mixed-methods approach captures both measurable shifts in literacy and the sociotechnical dynamics that shaped the Academy’s effectiveness. We address the following research questions:
\textbf{RQ1.} How effective was the \textit{AI Academy} in supporting instructors’ self-assessed development of generative AI literacies?
\textbf{RQ2.} What design and contextual factors influenced the Academy’s effectiveness?

The \textit{AI Academy} is not simply a training course to be evaluated for outcomes, but a case for understanding how instructors build generative AI literacies in relation to tools, policies, peer practices, and institutional supports. Our focus goes beyond measuring whether participants completed tutorials or improved skills with specific tools. Instead, we examine how design choices such as participatory co-construction of assessment instruments, collaborative reflection, and workflow redesign enable instructors to act as designers of their own responsible AI practices. Our contributions are threefold. First, we present the \textit{AI Academy} program model as a replicable approach to advancing instructor literacy. Second, we introduce a co-constructed survey instrument that both assesses and evolves with participants’ practices. Third, we provide design implications for AI tools that support faculty development through adaptive feedback, failure-aware exercises, scenario-based training, and personalized learning pathways. Together, these contributions position instructors as designers of responsible AI practice and contribute to the design of professional development platforms and ecosystems, such as those that provide adaptive assessment, safe spaces for ethical discussion, and opportunities for collaborative reflection and workflow redesign.

%% file: 2-relatedworks.tex
\section{Related Works}

\subsection{Teachers and Generative AI in Education}

Much early work on generative AI in education has focused on student use cases, from writing assistance to coding support and problem solving~\cite{kasneci2023chatgpt, zheng2023chatgpt}. These studies emphasize both opportunities (e.g., efficiency gains, rapid ideation, just-in-time feedback) and risks (e.g., overreliance, shallow learning, academic integrity concerns). In contrast, instructors---who mediate adoption and shape classroom practice---have received comparatively less systematic study. Unlike students, instructors must adapt pedagogical strategies, assessment practices, and professional roles to incorporate AI into teaching.

Research in HCI and the learning sciences demonstrates the potential of teacher-facing systems to enhance awareness, reflection, and instructional decision-making. For example, learning analytics dashboards improve real-time teacher awareness and promote data-informed reflection~\cite{holstein2019co, holstein2018student, Karumbaiah2023MultimodalAF, Alfredo2025TeamTeachingVizBC}. Professional development platforms and communities of practice highlight how digital systems can support ongoing growth and peer feedback~\cite{long2020ailiteracy,Alzoubi2024ACL}. In parallel, work on pedagogical agents and conversational support tools shows how systems scaffold not only student learning but also teacher adaptation, though these typically focus on K--12 or adaptive settings~\cite{apoki2022pedagogical, papoutsi2020pedagogical, peng2019personalized}. Importantly, these efforts largely predate generative AI and therefore do not address the distinct integration burden instructors face when adopting open-ended, general-purpose tools in higher education.

Our study responds to this gap by situating instructors as learners within the \textit{AI Academy}, an institutional program that surfaces obstacles, enabling conditions, and adoption patterns in generative AI for higher education.

\subsection{Generative AI Literacy Frameworks}

The concept of AI literacy builds on traditions of computational and digital literacy, emphasizing competencies such as understanding AI principles, critically evaluating outputs, and using AI responsibly~\cite{long2020ailiteracy, ng2021conceptualizing, robertson2025teaching}. These frameworks are increasingly applied in K--12 curricula and public initiatives, yet remain underexplored for higher education instructors ~\cite{tagare2025k}. Instructors face qualitatively different demands: aligning AI use with disciplinary epistemologies, balancing efficiency with integrity, and adapting teaching for new generations of learners.

Insights from child--AI interaction research further highlight AI literacy as a practice cultivated through design and reflection ~\cite{lin2020zhorai, newman2024want,su2022artificial}. For example, when children design AI agents, they develop agency and richer mental models of AI. Extending these findings to instructors underscores the importance of situated and reflexive literacy development, rather than static knowledge transmission.

Generative AI literacy extends AI literacy by emphasizing the skills and dispositions needed to engage critically with systems that generate content. It includes prompt design, evaluation of outputs, and awareness of ethical, legal, and professional issues such as bias, authorship, and intellectual property. These literacies are also socio-cultural, requiring reflection on how generative AI reshapes teaching, assessment, and disciplinary norms~\cite{Dewan_2025, ghimire2024generativeaieducationstudy, zastudil2023generativeaicomputingeducation}. Instructors act as both learners and mediators, and their approaches shape classroom norms, professional identity, and students’ responsible practices. Although student perspectives on generative AI in higher education have been studied~\cite{tang2025understandingstudentattitudesacceptability}, instructors remain comparatively understudied despite their central influence. Supporting their literacy therefore calls for professional development that integrates reflection, peer learning, and policy navigation, rather than focusing solely on technical tutorials.

\subsection{Measurement and Participatory Approaches}

While conceptual framings of AI literacy are advancing, measurement instruments tailored to professional educators remain underdeveloped. Existing surveys often capture broad public attitudes or student outcomes, leaving gaps in capturing instructors' evolving literacies in teaching contexts. Addressing this requires attention both to construct validity and to process.

Participatory approaches offer one promising path. Drawing on participatory research traditions, these methods emphasize co-construction, stakeholder involvement, and shared leadership. Prior studies show how participatory survey design surfaces locally meaningful priorities~\cite{kelley2019survey}, incorporates community voice, and identifies critical ``choice points'' for collaboration~\cite{vaughn2020participatory}. In policy contexts, participatory assessments have amplified marginalized voices and influenced decision-making~\cite{robb2002can}. Applying these traditions to AI literacy research ensures that measurement reflects instructors' lived realities, values, and concerns.

Together, these strands suggest three needs: (1) attending specifically to instructor adoption of generative AI, (2) reconceptualizing AI literacy for professional teaching contexts, and (3) using participatory methods to develop valid, context-sensitive instruments. Our study contributes to all three through the design and deployment of a participatory survey of higher education instructors in the \textit{AI Academy}.

%% file: 3-Method.tex
\section{Method}
\subsection{\textit{AI Academy} Design and Deployment} 
\subsubsection{Context and Participants}

The AI Academy was implemented as a faculty development initiative at a private university in the United States from January through May 2025. The program was designed to introduce instructors to generative AI concepts and practices, while encouraging critical reflection on pedagogy, ethics, and disciplinary norms. Participants were recruited through the university’s teaching and learning support networks, including faculty development newsletters, workshops, and direct outreach to departments. Participation was voluntary and open to all instructors. The study received IRB approval, and all participants provided informed consent for their data to be used for research purposes

\textit{Participants.} The cohort included instructors from multiple disciplines, including the humanities, social sciences, engineering, and professional programs. Participants varied widely in their teaching experience, ranging from early-career faculty to senior instructors, and in their prior familiarity with generative AI. This diversity provided opportunities for cross-disciplinary dialogue and highlighted distinct challenges and opportunities for AI integration across fields. 

\textit{Facilitators.} The Academy was coordinated by a group of ten facilitators who structured and guided the sessions. While they also held instructional roles at the university, their function within the Academy was limited to organizing content, moderating discussions, and providing resources. The team included individuals with expertise in pedagogy, educational technology, research, and program management. Responsibility for active engagement, exploration, and learning rested with the participants themselves. 

\textit{Research team within facilitators.} Two facilitators also served as the research team for this study. Their role extended beyond facilitation to include the integration of current scholarship into the program design and the iterative refinement of accompanying research instruments. Each session incorporated relevant research findings to strengthen both the Academy’s pedagogical structure and its survey measures. Throughout this paper, the pronoun \textit{``we''} refers exclusively to this research team. 

\subsubsection{Program Structure}

The program ran over the course of five sessions within an eight-week period. Each session was designed to balance conceptual learning, practical application, and reflective practice, ensuring that participants engaged with generative AI both as learners and as instructors shaping its role in higher education.

The format of the program blended whole-group workshops, hands-on challenges, and structured reflections. Participants explored AI literacy and ethics, practiced integrating AI into course design and assessment, and experimented with classroom applications. Sessions were intentionally interactive, creating space for participants to share perspectives, test tools, and adapt strategies to their disciplinary needs.

The learning activities combined short lectures to introduce key frameworks, small-group collaboration within disciplinary clusters, and guided practice with AI tools (see Appendix). Participants completed exercises such as prompt-engineering trials, syllabus redesign tasks, rubric creation, and classroom activity development. These were paired with reflective writing and peer discussions to deepen understanding and foster accountability.

At the end of the program, the Academy had a final session where participants shared the guidelines they created for their own disciplines and presented their findings to colleagues. This session gave participants the chance to pull together what they had learned, show practical examples, and discuss ideas for using AI in teaching in clear and responsible ways. This structure not only encouraged skill-building and experimentation but also emphasized collective responsibility for shaping ethical and effective AI use in teaching.

\begin{figure}[h]
    \centering
    \includegraphics[width=1\textwidth]{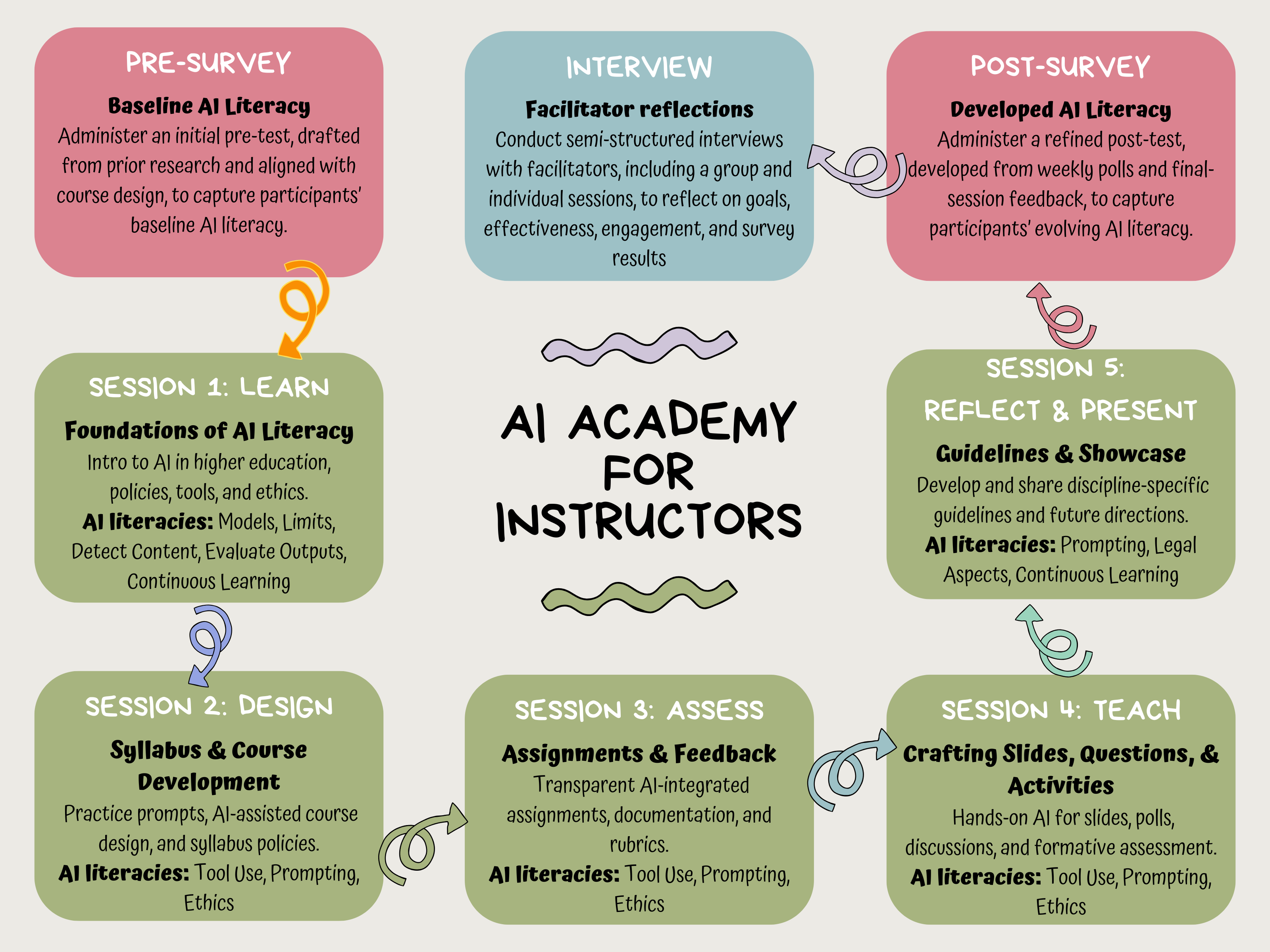} 
    \caption{Program structure of the AI Academy for Instructors. The flow includes a pre-survey, five instructional sessions, a post-survey, and facilitator interviews. Each session targeted specific dimensions of AI literacy, including knowledge, skills, ethical awareness, and reflective practice.}
    \label{fig:session1}
\end{figure}

\subsubsection{Learning Goals}

The Academy was designed to advance participants’ literacy in generative AI for teaching and learning. Specifically, the program emphasized three interconnected goals. First, participants were encouraged to develop a critical understanding of the responsible and ethical use of AI in educational contexts, including attention to academic integrity, equity, and transparency. Second, the Academy focused on the pedagogical integration of generative AI, supporting participants in identifying discipline-specific opportunities for incorporating AI into course design, assignments, and classroom activities. Third, participants practiced strategies for communicating about AI use with students and colleagues, with the aim of fostering shared expectations, informed dialogue, and clear guidelines for appropriate use. These goals reflected both the immediate needs of instructors and broader institutional priorities for navigating the evolving role of AI in higher education.

\subsubsection{Technology Choices}

The Academy incorporated a combination of institutionally supported platforms and widely available generative AI tools. Core activities were organized through a Canvas course shell, which provided a central space for session materials, asynchronous resources, and submission of reflective exercises. For synchronous engagement, sessions used Zoom for remote accessibility and in-person interactive technologies such as polling tools to model classroom practices. Participants experimented with a range of generative AI applications for teaching-related tasks (e.g., drafting syllabi, creating rubrics, designing assignments), focusing on tools approved for educational use and aligned with institutional policies. In addition, participants documented their learning through shared Google Docs and collaborative logs, which enabled reflection, transparency, and ongoing peer feedback. These choices balanced accessibility, institutional compliance, and opportunities for authentic exploration of AI tools in teaching practice.
\subsection{Pre- and Post-Survey Instrument Development and Participatory Research}

\subsubsection{Framework Alignment and Initial Design} 
The survey was developed through an iterative, collaborative process during the 2024--2025 winter term. 
The research team, in collaboration with Academy facilitators, drafted the initial instrument based on prior research on generative AI competence and AI literacy. 
The survey instrument was directly informed by Annapureddy et al.’s \textit{Generative AI Literacy: Twelve Defining Competencies} framework \cite{annapureddy2025generative}. Because the AI Academy targeted higher education instructors who already possessed baseline digital skills, we excluded the competency of \textit{basic AI literacy} and focused on the remaining 11 competencies. For each competency, the research team, in collaboration with facilitators, developed subscales consisting of 2–6 items adapted to the instructional and professional contexts of university teaching (e.g., course preparation, assessment design, student interaction). In addition, the competency “Knowledge of the contexts where generative AI is used” was further divided into four sub-dimensions: \textit{Policy and Academic Integrity}, \textit{Intellectual Growth}, \textit{Applications in Teaching and Learning}, and \textit{Communication with Students}.

\subsubsection{Pre-Test Deployment and Ongoing Feedback}
The initial version of the survey was administered as a pre-test in January 2025. 
Throughout the Academy, participants also completed weekly end-of-class polls, which provided real-time feedback on their concerns, challenges, and emerging questions. 
The research team reviewed these responses both to adjust program content and to identify areas where the survey might be refined.

\subsubsection{Participatory Refinement in the Final Session}
In the final session, participants revisited the pre-test questions and engaged in small-group discussions to suggest revisions. 
They identified which items were most relevant to their practice, which required clarification, and what new questions should be added in response to both their experiences 
during the Academy and the rapidly evolving landscape of generative AI. 
Participants engaged in a classroom activity titled *“Becoming AI Literacy Leaders in Your Discipline.”* Each group received a subset of ten survey items representing different AI literacy dimensions. 
Over 15 minutes, groups were asked to (1) reflect on which items resonated most or least in their discipline, (2) critique the framework by suggesting additions, removals, or clarifications, 
and (3) envision how they would guide colleagues or students as AI literacy leaders. 
Following the discussion, groups wrote short reflections in their shared workspace and presented one takeaway to the full cohort in the final 20 minutes.

\subsubsection{Instrument Revision and Post-Test}
During the Academy, participants provided feedback on the clarity and redundancy of survey items. 
Based on this feedback, we revised or removed items that were overlapping or difficult to interpret, resulting in a more streamlined post-survey instrument. 
The final version contained 9 subscales, reflecting competencies that participants viewed as both meaningful and distinctive for their teaching practice, 
and was deployed as the post-test in May 2025.

\subsection{Data Collection and Analysis}

To evaluate the Academy’s impact and capture participants’ experiences, we employed a mixed-methods design combining quantitative surveys, open-ended feedback, and facilitator interviews. 
\subsubsection{Pre Post Survey from Participants}
To evaluate changes in participants’ AI literacy, we administered surveys at the beginning and end of the AI Academy. A total of 25 participants completed the pre-survey and 16 completed the post-survey. The pre-survey instrument included 43 items; in the post-survey, 36 items were retained verbatim, 5 were revised for clarity or specificity, and 2 were removed, resulting in 41 items in total. Items were rated on a 5-point Likert scale (1 = strongly disagree to 5 = strongly agree).

Items were organized into subscales, each consisting of 2--6 items designed to capture a specific dimension of AI literacy 
(e.g., understanding AI concepts, confidence in AI use, ethical awareness). For each participant, we computed subscale scores 
as the mean of their constituent items. Because pre- and post-surveys were not fully matched (25 pre, 16 post), we treated 
them as independent samples. For each subscale, we compared pre- and post-survey means using Welch’s independent-samples 
$t$-tests, which are robust to unequal sample sizes and variances. As shown in Table~\ref{tab:alpha_results_full}, 
internal consistency of each subscale was assessed using Cronbach’s $\alpha$. These results are exploratory and should not 
be considered sufficient for a validated survey. Most subscales improved from exceeding the 0.70 threshold to exceeding the 
0.80 threshold, suggesting stronger reliability at post-survey, though weaker consistency was observed for 
\textit{Knowledge of capacity \& limitations}, \textit{Assess AI output}, and \textit{Context knowledge -- Policy \& Academic Integrity}.

While the survey was used to measure AI literacy outcomes, it is not a psychologically validated instrument. Establishing its validity and reliability would require further research and is not the contribution of this paper. Our focus is on the participatory process that shaped the survey items and highlighted constructs of AI literacy relevant to higher education instructors.

\subsubsection{Open-Ended Feedback from Participants}

Due to the variability in faculty schedules and communication preferences, participants were invited to provide open-ended feedback through multiple channels (optional): (1) weekly reflection “homework” prompts submitted in the Canvas course shell, (2) short post-session surveys administered after every session, (3) written notes recorded during activities, and (4) contributions to a collaborative Google Doc “learning log”. Twenty-one participants contributed to the learning log. In addition, full-session recordings were obtained, which captured whole-class activity, including whole-group sharing sessions. Aggregating across these sources ensured that diverse perspectives were captured, including those from less vocal participants.

All available qualitative responses were analyzed using inductive thematic analysis \cite{braun2006using}. The research team carried out coding, iteratively refining categories and documenting analytic decisions through reflexive memos. Interpretations were collaboratively reviewed within the team to strengthen validity. Common themes included excitement about teaching possibilities, confusion around AI policies and tool limitations, and concerns about workload.

\subsubsection{Organizer Interviews}

To complement participant perspectives, we conducted semi-structured interviews with the Academy facilitators ($N=6$). They are referred to as F1 to F6 in the manuscript. These included one group interview at the conclusion of the program and individual interviews with each facilitator. The protocol included reflections on program goals, session effectiveness, and participant engagement. Organizers were also presented with aggregated results from the pre–post surveys and asked to comment on observed changes. All interviews were audio-recorded, transcribed verbatim, and thematically coded using the same iterative process as participant feedback. Thematic saturation was reached after coding all transcripts.

%% file: 4-Findings.tex
\section{Findings}\label{sec:findings}

\subsection{RQ1: Usefulness of the \textit{AI Academy} for Instructor Learning}\label{sec:rq1}

\begin{figure}[h]
    \centering
    \includegraphics[width=0.9\textwidth]{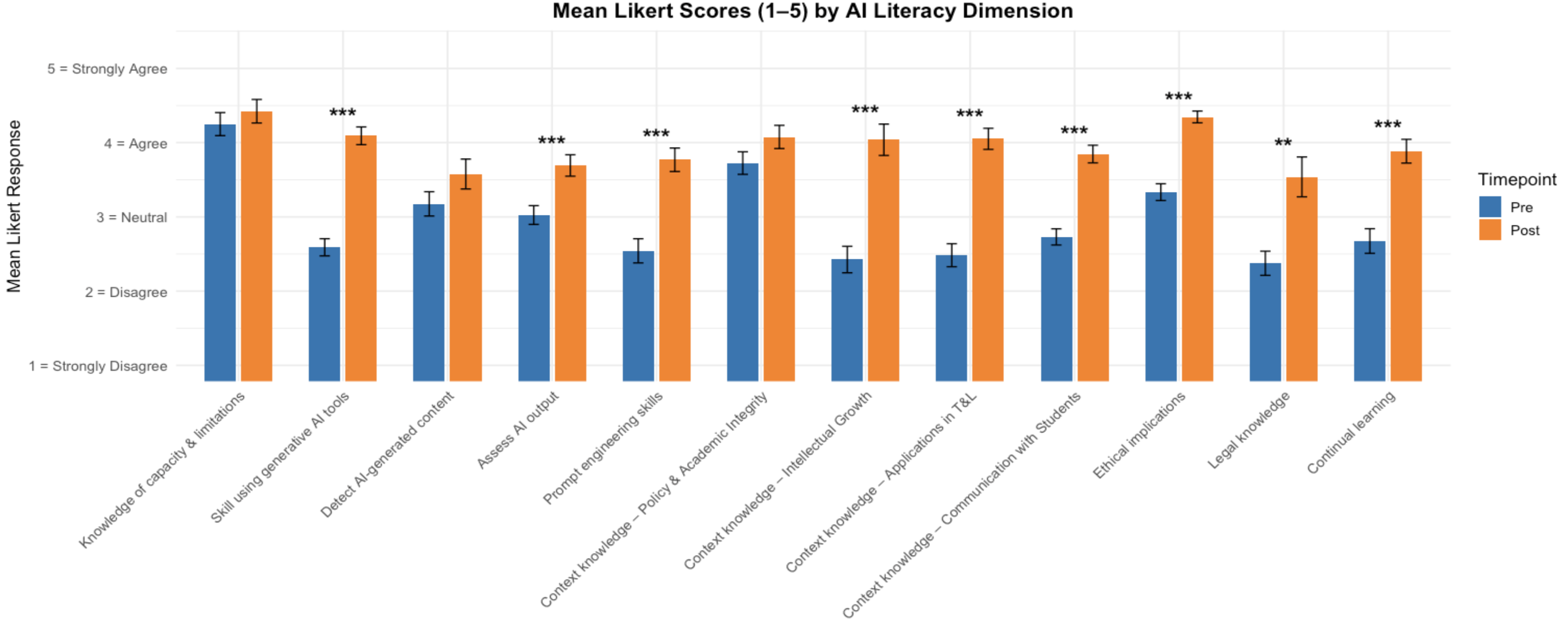} 
    \caption{Pre–post survey results for faculty participants across AI-literacy dimensions. Significant gains ($^{***}p \leq 0.001$) were observed in assessing AI outputs, most subdimension of contextual knowledge, continual learning, ethical implications, prompt engineering, and skills in using generative-AI tools. Moderate improvement ($^{**}0.001 < p \leq 0.01$) was found in legal knowledge. No significant change was detected for detecting AI-generated content, knowledge of AI capacity \& limitations and context knowledge -- policy \& academic integrity. Error bars show standard error of the mean.}

    \label{prepost}
\end{figure}

We examine the usefulness of the \textit{AI Academy} for instructor learning using (1) group discussions where instructors critiqued and refined generative AI literacy items, and (2) pre/post survey data. Together, these sources show significant gains across literacy dimensions (Figure~\ref{prepost}) and make visible how instructors appropriated and reshaped the literacy items in relation to their own disciplines. The activity enabled faculty to engage deeply with generative AI literacy not only as learners but as designers of the very instruments used to capture it, including revising wording and tone, removing items not useful for teaching practice, etc.

\subsubsection{Pre and Post Survey Results} 
As see in Figure \ref{prepost}, pre–post survey results show \emph{significant gains} in nearly all AI-literacy dimensions among \emph{faculty participants}. Paired tests indicated highly significant improvements ($^{***}$, $p \le 0.001$) in assessing AI outputs, contextual knowledge of AI, continual learning about AI, ethical implications, prompt engineering skills, and skill using generative-AI tools. A moderate improvement ($^{**}$, $0.001 < p \le 0.01$) was observed for legal knowledge. Two dimensions did not improve significantly: detecting AI-generated content and knowledge of AI capacity and limitations (Figure~\ref{prepost}). Post-hoc power analyses indicated that most dimensions had large effect sizes (Cohen’s d > 0.8) with statistical power exceeding 0.90, suggesting robust improvements from pre to post. Exceptions were Knowledge of capacity \& limitations (d = 0.19, power = 0.11) and Detect AI-generated content (d = 0.38, power = 0.33), which showed small effects and insufficient power, indicating little measurable change in these areas. Overall, the \textit{AI Academy} effectively advanced participants’ skills across most dimensions, while AI detection and understanding of limitations remained difficult areas likely requiring deeper or longer-term engagement.



\subsubsection{Pre--Post Survey Item Major Changes} The pre-survey contained 41 items. In the post-survey, 27 items were unchanged, 9 were revised (Q6, Q7, Q20, Q21, Q28, Q31, Q35, Q37, Q38), and 5 were removed/merged with others (Q1, Q2, Q18, Q19, Q39), resulting in a 36-item post instrument.

In \textit{Skills to Use Generative AI Tools}, the post-survey revised wording to align with educational goals and added concrete examples (e.g., rubric development), making items clearer and more practical.  

The \textit{Programming and Fine-tuning} category, which included awareness of concepts and tools, was removed entirely after instructors noted it was not relevant to their teaching practice.  

In \textit{Ethical Implications}, one item was broadened to include creativity and other cognitive skills, and another was replaced with a reflection on the purpose of higher education and the rigor of student learning, expanding the scope beyond classroom-level ethics.  

In \textit{Legal Aspects}, separate items on copyright and publishing were merged into a single, streamlined item on legal and copyright implications of using and sharing AI-generated teaching materials.  

Beyond these major revisions, most other dimensions (e.g., Knowledge of Capacity and Limitations, Detecting and Assessing AI Outputs, Prompting Skills, Applications in Teaching and Learning, Intellectual Growth, Communication with Students, and Lifelong Learning) were retained with only minor wording adjustments to improve clarity, responsibility, or specificity. These refinements reflect instructors’ feedback for clearer phrasing while maintaining major consistency across the pre- and post-surveys.  

\subsubsection{Participant Discussion in Survey Refinement} In the final session, participants took part in ``Becoming AI Literacy Leaders in Your Discipline.'' Drawing on knowledge gained in the Academy, each group reviewed ten survey items, discussed which resonated in their field, critiqued wording, and identified priorities for guiding colleagues and students. Their comments were later taken up in facilitator discussions and incorporated by the research team into revisions of the survey instrument.

\textbf{Group 1} (Q1-Q10) emphasized ethics and instructional practice. One participant noted that “\textit{law students need to be extremely mindful about the outputs of these systems and what additional steps their ethical obligations require},” underscoring accuracy and bias in professional education. The group recommended revising assignment-design items to “\textit{preserve the educational goals of the assignments while making decisions about student use of AI}” and adding an item on supporting responsible student use. Knowledge-level items on AI models were described as less relevant, “\textit{too abstract compared to what we actually need in teaching}.”

\textbf{Group 2} (Q11-Q19) centered on assessing AI outputs and prompting skills. Participants highlighted the importance of subject-matter expertise when evaluating inaccuracies, noting that AI-generated code “\textit{didn’t work… but I had to know to ask for that}.” They emphasized the need to teach students how to diagnose and reprompt effectively, since “\textit{knowing how to, when it gives you something wrong… how to keep prompting it}” was seen as essential for making AI a useful tool rather than “\textit{just magic}.” Their discussion also connected detection and assessment to broader ethical concerns, asking “\textit{what are we losing? What does this mean for thinking?}” and stressing the need for space to wrestle with both skills and ethics in faculty development.

\textbf{Group 3} (Q32-Q41)  converged on several points: they emphasized contextual variability and cautioned against assuming faculty can outpace students, “\textit{students are going to be ahead of the game…they would be unimpressed by many faculty’s knowledge}”; They expressed a preference for shared institutional resources over placing responsibility on individuals. At the same time, they drew distinctions across domains: in teaching, they emphasized the need for explicit ethical guidelines and rules, while in research, they approached AI both as a social reality that shapes practices and as a technical object to be critically examined. They also flagged gaps in the instrument and program, insufficient technical coverage (e.g., programming, fine-tuning) and too little attention to higher education’s role in sustaining rigor. 

\textbf{Group 4 \& 5} (Q20--Q31) These groups emphasized applications in teaching and learning, especially the importance of knowing available tools to guide students in line with industry expectations, drawing on hands-on experience from the workshop. They called for stronger attention to applied tool use and noted marked disciplinary differences in the value of mastering specific tools. Building on these themes, Group 5 requested clarification of who ``transparency'' applies to (faculty, students, or both), and identified gaps concerning AI's influence on creativity and cognitive development, as well as the long-term balance between analog and digital learning. They further critiqued policy and integrity items in light of unreliable detection tools, noting that ``\emph{trying to limit AI use at present merely rewards those who are the best at hiding AI usage}.'' For Q20, they suggested revising the item ``I provide clear explanations of what constitutes fair and dishonest use of generative AI tools in my class'' to: ``I clearly define acceptable and unacceptable uses of generative AI in my class to uphold academic integrity, recognizing the limitations of current AI detection tools.'' They foregrounded foundational and ethical dimensions (critical thinking, inequalities, copyright, privacy, adaptability) and cautioned that, because AI is rapidly evolving and individuals’ self-assessments may be unreliable, survey language should avoid overstating certainty. For example, they recommended revising ``\emph{I understand}'' to ``\emph{I am beginning to develop an understanding}.''


\textbf{Summary of RQ1.} Pre–post results showed significant gains in six of nine Generative AI-literacy dimensions, a moderate gain in legal knowledge, and no measurable change in detecting AI-generated content or knowledge of AI capacity and limitations. Participants’ critiques and revisions produced a refined 36-item instrument that was both contextually grounded and co-constructed. Taken together, these findings imply that the \textit{AI Academy} advanced faculty self-assessed learning while also positioning instructors as designers of responsible AI practice through the very instruments used to assess it.

\subsection{RQ2: Practices and Challenges in Advancing Teachers’ Generative AI Literacy}\label{sec:rq2}  
Analysis of \textit{facilitator interviews} and \textit{participant learning logs} examines how the \textit{AI Academy} supported instructors in designing their own approaches to generative AI in teaching and learning. The data highlight how participants experimented with tools, rethought assessments and transparency, and grappled with ethical and pedagogical tensions, showing both the possibilities and challenges of positioning faculty as AI designers in their disciplines.


\subsubsection{Facilitator Interview on AI Academy Process and Outcomes} 
Across the six facilitator interviews, three broad themes emerged regarding what shaped the effectiveness of the AI Academy and where improvements could be made. These super-themes integrate points of agreement and highlight ongoing tensions in course design and delivery.  

\textbf{From Literacy to Fluency: Building Critical Awareness of AI}  
Facilitators emphasized that effectiveness was not only about acquiring skills but also about recalibrating understanding of AI’s limits and possibilities. Participants moved from overconfidence to what F1 described as “conscious incompetence,” realizing how much they did not yet know: \textit{“The post actually being very close to the pre shows that, because participants became more consciously literate, they realized how much more they need to learn”} (F1). Prompt engineering was repeatedly described as a core literacy, requiring modeling and guided practice. As F1 explained, \textit{“Prompt engineering is one of the biggest and most important core literacies.”} Sequencing was also highlighted, anchoring first in pedagogy, then in tools, to avoid technology-first thinking. F1 articulated this view explicitly: \textit{“The technology should be last, and pedagogy should always come first.”}  

Many facilitators stressed that literacy must also extend to understanding the limits of AI, including the contested space of detection. While F3 dismissed detection tools as unreliable, noting, \textit{“There are no tools out there that can accurately say this is AI-generated content, and as a university, we are not moving in that direction,”} F2 suggested that detection could still be taught as literacy rather than enforcement: \textit{“Detection should be taught as literacy rather than enforcement.”} Finally, facilitators agreed that the academy should evolve from general literacy toward domain-specific fluency that links AI directly to teaching and research practices. As F1 put it, \textit{“We need to move from general AI literacy to domain-specific AI fluency, which means being efficient, effective, and ethical within your own discipline.”}  

\textbf{Designing with Peers: Collaboration and Institutional Ecosystems}  
The academy’s effectiveness was amplified by its social and institutional context. Peer interaction, such as comparing artifacts and sharing prompts, helped participants learn from one another. F5 emphasized the role of group activities, explaining, \textit{“When people grouped together to figure out the activities and see each other’s results, it increased everybody’s skill level and what they learned from the session.”} At the same time, disciplinary differences shaped engagement. F1 observed that \textit{“Humanities faculty are really getting snagged on ethical and philosophical issues,”} but facilitators also noted that hands-on use could help shift such skepticism into more informed judgment.  

Leadership decisions and institutional support further shaped engagement. Provisioning access created equity in tool use, as F1 reflected: \textit{“When our institution turned on Gemini for everybody, that flattened the access curve.”} Strong recruitment efforts also generated momentum, with later cohorts joining based on reputation and peer word-of-mouth. Facilitators additionally stressed the importance of scaling through distributed support roles. F5 proposed pairing faculty with graduate students and sustaining engagement through coaching: \textit{“If we assign participants to work with a graduate student and continue through AI coaching, they would have someone to reach out to and continue their engagement after the academy.”} Similarly, others suggested embedding faculty fellows or student AI coaches to maintain community impact.  

\textbf{Making Learning Visible: Evidence, Artifacts, and Sustainable Impact}  
Facilitators highlighted the need to both evidence and extend learning beyond the academy. Tangible outputs, such as capstone projects or a digital gallery, were proposed as a way to demonstrate impact to participants and leadership alike. F5 argued, \textit{“It would be a good idea to have a capstone project developed across all five classes, with the final project displayed on a gallery or website.”} Supporting infrastructure such as high-quality HyFlex recordings, integrated journals, and digital badges also helped track engagement and create accountability. HyFlex is a hybrid–flexible course design model that lets students choose to participate in person, online in real time, or asynchronously, with all formats leading to the same learning outcomes.  F1 described these tools as integral to sustaining progress: \textit{“We had Canvas, learning analytics data, and digital badges, which helped track and nudge progress.”}  

Finally, facilitators noted that a successful outcome was not just skill acquisition but motivating participants to continue learning and experimenting with AI in their own contexts. As F1 summarized, \textit{“Continual learning is what matters. Now it looks like they are motivated to learn more about what they don’t know.”} Across these accounts, effectiveness was tied not only to what participants learned during the academy but also to the infrastructure and artifacts that sustained their growth beyond the sessions.  

\subsubsection{Open Ended Feedback from Participants} Analysis of participant learning logs provides complementary insights into how the Academy shaped faculty engagement. Rather than offering direct evaluations, the logs revealed areas where the Academy deepened conceptual understanding, encouraged experimentation, and surfaced ethical and pedagogical concerns. 

\textbf{Experimentation in Practice: Clarifying Concepts and Trying Tools
} Participants described how the Academy improved their understanding of AI fundamentals. One participant highlighted the role of structured prompting: \textit{``prompt engineering – learned a lot about this, essential for all the other work of the academy''} (P2). Another reflected on newfound conceptual clarity: \textit{``today for the first time I understood what exactly the term `generative AI’ means and how it differs from an LLM''} (P7). Such moments of clarity gave instructors confidence to engage with AI from a more informed stance. At the same time, participants noted when coverage was too general: \textit{``lots of information was given. I feel more informed generally but not specifically. Still lots to learn and practice with to feel more comfortable''} (P1). These tensions highlight how literacy-building was effective, but also point to a need for more discipline-specific tailoring. The Academy was most impactful when participants experimented directly with their own teaching contexts. One reported, \textit{``I am now using AI more often to create assignments, quizzes, discussion questions''} (P1). Another emphasized efficiency: \textit{``Pria proved to be a useful tool for structuring and refining course materials, offering efficiency in content organization and iterative syllabus development''} (P14). 

\textbf{Ethics in the Classroom: Rethinking Assessment and Integrity} A defining feature of the Academy was how it prompted instructors to grapple with ethical and pedagogical dilemmas. One participant warned of \textit{``the dystopian future of AI’s producing all of student work – the loss of writing as a skill and a training/learning tool, the harms to literacy and science''} (P2). Another noted, \textit{``I have concerns about ethical use of it by students''} (P1). As one noted, the value lay in “\textit{having space to talk about the ethical questions and then understanding how to use it}. 

\begin{quote}
    \textit{My thoughts: In my own course, I found it very difficult (and more importantly time-consuming) to try to police these types of activities. I think it is more effective to change the nature of assessments to be able to distinguish students’ learning from them relying too heavily on AI for getting through the course. I do not have all the answers yet, but achieving this has meant I needed to adopt a radically different mindset about how my assessments work, what academic integrity means in face of AI, how AI could be used more as a tool to enhance learning rather than something to fight against, etc. I am still very much in the process of figuring this all out.} (P16)
\end{quote}

This extended reflection illustrates how the Academy spurred instructors to reconceptualize AI not simply as a threat, but as a catalyst for reassessing integrity and learning outcomes. 

\textbf{Productive Struggles: Learning Through AI’s Limitations
} Participants frequently encountered limitations with AI systems, but these struggles became moments of critical learning. One observed, \textit{``it did create the PPT but it wasn’t visually appealing''} (P11), while another found, \textit{``ChatGPT’s comparison included a number of mischaracterizations… I found both answers rather superficial and to lack citations and detailed examples and explanations''} (P13). Others described the importance of iteration: \textit{``this example clearly shows how important using correct prompts in generative AI really is. This was a great learning experience for me''} (P9). Participants also reflected on the balance between usefulness and over-reliance: \textit{``good for brainstorming syllabus ideas and assignments (less so for making syllabi wholesale)''} (P2). These frustrations were not failures but productive struggles that sharpened participants’ critical stance toward AI outputs. As one also put it, \textit{``I want to leverage what it can offer but I am nervous about losing student creativity and innovation. I am also afraid of what I or others may lose by skipping the `struggling’ stage''} (P11).

\textbf{Elicitation of Reflections on Faculty–Student Asymmetries.} A final factor shaping the Academy’s effectiveness was how it made visible the asymmetries between faculty and students in their use of AI. Through the Academy, many instructors became more aware of how their students could use AI in both beneficial and problematic ways. This awareness surfaced new tensions, as participants grappled with questions of fairness, transparency, and professional identity. One instructor reflected, \textit{``I’m much more attuned to the ways that students can evade learning through AI''} (P2), while another noted concerns about ethical boundaries: \textit{``I have concerns about ethical use of it by students''} (P1). At the same time, some participants recognized that AI might serve as a tool for creativity and learning if integrated thoughtfully.  

These tensions extended beyond student behavior to issues of disclosure and credibility in faculty practices. As one participant asked, \textit{``students report that not all of their professors have class policies addressing generative AI''} (P16), pointing to uneven institutional guidance. Another reflected on the dilemma of whether to disclose their own AI use:  

\begin{quote}
    \textit{How different is using AI than having access to a bank of exam/quiz questions from a publisher, etc. On one hand, I think disclosure may be warranted given we are paying so much attention to how students use AI right now in classes. But on the other hand, does that reveal too much going on behind the curtain? … Emphasis on the use of AI by teachers takes away the decades of schooling/experience that also goes into teaching. We don’t want to give the impression we are just pushing buttons to get our teaching materials when it is so much more than that – always.} (P16)
\end{quote}

\textbf{Summary of RQ2.}  
Facilitator interviews and participant reflections show that the \textit{AI Academy} was most effective when it clarified concepts, encouraged experimentation, and created space for ethical and pedagogical reflection. These activities helped instructors redesign their classroom practices around assessment, transparency, and professional identity, positioning them as active designers of responsible AI use in teaching with potential to shape broader norms in higher education.

%% file: 5-Discussion.tex
\section{Discussion}

\subsection{Generative AI Literacy Survey: Gains, Gaps, and Participatory Research}
Pre–post results show that the \textit{AI Academy} advanced instructors’ generative AI literacies, with significant gains across six of nine dimensions and a moderate gain in legal knowledge. The participatory survey design was central to this outcome: faculty critiqued, refined, and restructured items, yielding a streamlined 36-item post instrument aligned with disciplinary contexts. Revisions included removing the \textit{Programming and Fine-tuning} category, merging legal items, and clarifying tool-use skills with teaching examples (e.g., rubric design).  

The two dimensions that did not improve significantly, \textit{Detecting AI-generated content} and \textit{Knowledge of capacity \& limitations}, were largely unchanged, suggesting that the absence of gains reflects enduring pedagogical challenges rather than flaws in measurement. Facilitators described this as “conscious incompetence,” where participants became more aware of AI’s limits and uncertainties rather than more confident. In this sense, non-significant results may still represent meaningful learning not fully captured by pre–post statistics. The \textit{Context knowledge – Policy \& Academic Integrity} scale also showed no significant gain. Unlike the other two, these items were revised to foreground enforcement practices, aligning with qualitative findings that participants struggled to move from policy awareness to classroom implementation amid unreliable detection tools and evolving guidelines.  

Reliability patterns further underscore the instrument’s value as both measure and developmental artifact. Cronbach’s $\alpha$ improved across most scales (Table~\ref{tab:alpha_results_full}); for example, \textit{Continual learning} rose from poor (.56) to excellent (.90), while \textit{Knowledge of capacity \& limitations} improved from unacceptable (.10) to questionable (.67). These shifts suggest more consistent interpretations after collective engagement. Scales with fewer than three items should be examined and possibly enriched. Future validation with larger samples and advanced psychometric approaches such as confirmatory factor analysis and item response theory is needed \cite{devellis2017scale,boone2016item}.

\subsection{From Literacy to Fluency: Positioning Instructors as Designers of Responsible AI Practice}
RQ1 and RQ2 together show that the \textit{AI Academy} supported instructors in moving from literacy to fluency with generative AI. Pre–post surveys demonstrated significant gains in assessing outputs, contextual knowledge, continual learning, ethical implications, prompt engineering, and tool use, while group discussions showed how instructors critiqued and reshaped literacy items in relation to their own disciplines. This co-construction positioned faculty not only as learners but also as designers of the very instruments used to capture learning.  Facilitator interviews emphasized that effectiveness was not only about acquiring skills but also about recalibrating understanding of AI’s limits and possibilities. Prompt engineering was repeatedly described as a core literacy, requiring modeling and guided practice, and sequencing was highlighted to ensure pedagogy guided tool use. Participant logs further showed experimentation with assignments, quizzes, and course design, along with productive struggles when tools were inaccurate or limited. The Academy also created space for ethical and pedagogical reflection, where participants moved from policing misuse toward redesigning assessments and transparency.  Taken together, these findings show that the Academy clarified concepts, strengthened core literacies, encouraged experimentation, and created opportunities for ethical and pedagogical reflection, while also highlighting persistent challenges in detection, capacity and limitations, and policy and academic integrity.

\subsection{Design Implications for AI Tools to Support Faculty Development}

\textbf{Adaptive Feedback for Prompting and Assessing Outputs.}  
Prompt engineering and assessing AI outputs were among the strongest survey gains. AI tools could build on this by providing adaptive feedback: generating multiple responses to the same prompt, highlighting differences, and explaining why certain formulations lead to more aligned results. Similar to intelligent tutoring systems, adaptive feedback can strengthen skill transfer and confidence in real-world tasks \cite{anderson1995cognitive, holstein2019student}. For assessment, tools could surface weaknesses in outputs, such as unsupported claims, helping faculty practice systematic critique rather than relying on intuition.  

\textbf{Failure-Aware Exercises for Capacity and Detection.}  
Knowledge of AI capacity and limitations, along with detection of AI-generated content, showed no measurable change. AI tools can address this gap by generating deliberate failure cases, hallucinated references, biased outputs, or overly superficial explanations, and asking instructors to identify flaws. Confidence markers and error overlays can reinforce this literacy by making blind spots visible. Such failure-based learning echoes work on productive struggle as a pathway to deeper understanding \cite{kapur2008productive, wang2024blindspots}.  

\textbf{Scenario Generation for Policy and Ethical Implications.}  
Context knowledge of policy and integrity, along with ethical implications, remained challenging. AI tools could create branching scenarios where faculty practice responding to suspected misuse or disclosure dilemmas, exploring outcomes of different choices. Scenario-based training is widely used in professional development to address ethical and policy challenges \cite{narayanan2020aiethics, lee2022scaffolding}. Tools could also generate draft policy language tailored to disciplinary norms, giving faculty a concrete starting point for classroom integration.  

\textbf{Personalized Pathways for Continual Learning and Communication.}  
Continual learning and communication with students require sustained support. AI tools could personalize learning trajectories: micro-lessons for novices, discipline-specific case libraries for advanced users, and progress dashboards to make growth visible. Collaborative galleries of prompts, assignments, and integrity statements would further allow peer learning. These approaches align with research showing that differentiated scaffolding and communities of practice strengthen professional development \cite{desimone2009improving, koh2023genai}.

%% file: 6-Conclusion.tex
\section{Conclusions and Limitations}

This paper introduced the \textit{AI Academy}, a faculty development program designed to advance instructors’ generative AI literacy. Using pre/post surveys, learning logs, group discussions, and facilitator interviews, we observed gains across most literacy dimensions alongside persistent challenges, including uncertainty about AI’s limits and learning moments around detection as literacy rather than skill mastery. The Academy functioned not only as a training program but also as a participatory research space, where instructors critiqued and refined the survey instrument and engaged in peer reflection on ethics, credibility, and professional identity. These findings show that professional development is most effective when treated as an evolving ecosystem—adapting to new tools, providing authentic practice environments, surfacing AI’s fallibility, and creating safe spaces for ethical discussion. In this way, the work moves from literacy to fluency, positioning instructors not only as learners but as designers of responsible AI practices within their classrooms and institutions.

Our study has several limitations. First, it was conducted at a single private U.S. university, which may limit transferability to other institutional contexts or international settings. Second, the cohort was modest in size, constraining both statistical power and the diversity of disciplinary perspectives. Third, the pre/post survey, though co-constructed, relied on self-reported literacy rather than observed classroom practice; future work should triangulate these measures with classroom observations or student outcomes. Finally, given the rapid evolution of generative AI, specific tools and concerns discussed during the Academy may shift quickly. These limitations underscore the need for iterative, scalable models of professional development and for technical systems that can adapt dynamically as AI technologies and educational practices continue to evolve.

%% file: 7-appendix.tex
\clearpage

\appendix
\section{Survey Items (Pre vs. Post)}
\renewcommand{\arraystretch}{1.25}
\begin{table}[ht]
\centering
\scriptsize
\begin{tabular}{|p{2.7cm}|p{0.5cm}|p{6cm}|p{6cm}|}
\hline
\textbf{Category} & \textbf{ID} & \textbf{Pre} & \textbf{Post} \\
\hline
Knowledge of generative AI models & Q1 &
I understand that generative AI is a statistical model trained on large datasets and can produce various types of output. &
Deleted \\
\hline
Knowledge of generative AI models & Q2 &
I am aware that generative AI is increasingly integrated into various applications, such as search engines. &
Deleted \\
\hline
Knowledge of generative AI models & Q3 &
I can explain the differences between generative AI and other types of AI (e.g., rule-based or predictive AI). &
I can explain the differences between generative AI and other types of AI (e.g., rule-based or predictive AI). \\
\hline
Knowledge of the capacity and limitations & Q4 &
I understand that generative AI tools can produce creative outputs but may also generate incorrect or misleading information. &
I understand that generative AI tools can produce creative outputs but may also generate incorrect or misleading information. \\
\hline
Knowledge of the capacity and limitations & Q5 &
I can explain how biases in training data influence generative AI outputs. &
I can explain how biases in training data influence generative AI outputs. \\
\hline
Skill to use generative AI tools & Q6 &
I know how to use generative AI tools to develop effective course designs or structures. &
I know how to use generative AI tools to develop effective course designs or structures that achieve the education goal. \\
\hline
Skill to use generative AI tools & Q7 &
I know how to create assignments with the help of generative AI tools. &
I know how to create assignments with the help of generative AI tools, such as supporting the creation of rubrics. \\
\hline
Skill to use generative AI tools & Q8 &
I know how to employ generative AI tools to brainstorm and refine course topics. &
I know how to employ generative AI tools to brainstorm and refine course topics. \\
\hline
Skill to use generative AI tools & Q9 &
I know how to use generative AI tools to enhance the quality of student feedback. &
I know how to use generative AI tools to enhance the quality of student feedback. \\
\hline
Skill to use generative AI tools & Q10 &
I know how to use generative AI tools to cater to diverse student groups and individual learning requirements. &
I know how to use generative AI tools to cater to diverse student groups and individual learning requirements. \\
\hline
Ability to detect AI-generated content & Q11 &
I know how to distinguish between AI-generated and human-created content in written assignments. &
I know how to distinguish between AI-generated and human-created content in written assignments. \\
\hline
Ability to detect AI-generated content & Q12 &
I know how to identify when images or visuals are created using generative AI tools. &
I know how to identify when images or visuals are created using generative AI tools. \\
\hline
Ability to assess AI outputs & Q13 &
I know how to identify biases in outputs from generative AI tools. &
I know how to identify biases in outputs from generative AI tools. \\
\hline
Ability to assess AI outputs & Q14 &
I know how to recognize when outputs from generative AI tools are inaccurate or incomplete. &
I know how to recognize when outputs from generative AI tools are inaccurate or incomplete. \\
\hline
Skill in prompting generative AI tools & Q15 &
I know how to design and refine prompts to achieve high-quality and relevant outputs. &
I know how to design and refine prompts to achieve high-quality and relevant outputs. \\
\hline
Skill in prompting generative AI tools & Q16 &
I am aware of technical concepts for advanced prompt techniques (RAG, few-shot, chain-of-thought). &
I am aware of technical concepts for advanced prompt techniques (RAG, few-shot, chain-of-thought). \\
\hline
Skill in prompting generative AI tools & Q17 &
I know how to use advanced prompt techniques such as examples and step-by-step instructions. &
I know how to use advanced prompt techniques such as examples and step-by-step instructions. \\
\hline
Ability to program and fine-tune & Q18 &
I am aware of the technical concepts for fine-tuning generative AI models. &
Deleted \\
\hline
Ability to program and fine-tune & Q19 &
I know how to use data, platforms, and tools used to fine-tune generative AI models. &
Deleted \\
\hline
\end{tabular}
\caption{Pre and Post Survey Items for Gen AI Literacy (Knowledge, Skills, Detection, Prompting).}
\label{tab:prepost1}
\end{table}

\begin{table}[ht]
\centering
\scriptsize
\begin{tabular}{|p{2.6cm}|p{0.5cm}|p{6cm}|p{6cm}|}
\hline
\textbf{Category} & \textbf{ID} & \textbf{Pre} & \textbf{Post} \\
\hline
Context knowledge -- Policy and Academic Integrity & Q20 &
I provide clear explanations of what constitutes fair and dishonest use of generative AI tools in my class. &
I clearly define acceptable and unacceptable uses of generative AI in my class to uphold academic integrity, recognizing the limitations of current AI detection tools. \\
\hline
Context knowledge -- Policy and Academic Integrity & Q21 &
I follow the university’s guidelines for generative AI and academic integrity. &
I follow the university’s guidelines for generative AI and academic integrity, for example how to address suspected inappropriate use of generative AI in my classroom. \\
\hline
Context knowledge -- Intellectual Growth & Q22 &
I know how to design and implement new teaching practices that integrate generative AI for my professional growth. &
I know how to design and implement new teaching practices that integrate generative AI for my professional growth. \\
\hline
Context knowledge -- Intellectual Growth & Q23 &
I know how to create innovative learning activities using generative AI to enhance student growth. &
I know how to create innovative learning activities using generative AI to enhance student growth. \\
\hline
Context knowledge -- Applications in Teaching and Learning & Q24 &
I am aware of discipline-specific applications of generative AI tools that are relevant to my discipline, such as writing assistance. &
I am aware of discipline-specific applications of generative AI tools that are relevant to my discipline, such as writing assistance. \\
\hline
Context knowledge -- Applications in Teaching and Learning & Q25 &
I can describe how generative AI tools impact broader educational goals like personalized learning. &
I can describe how generative AI tools impact broader educational goals like personalized learning. \\
\hline
Context knowledge -- Applications in Teaching and Learning & Q26 &
I know how to use generative AI tools to address diverse learning needs using multimodal approaches. &
I know how to use generative AI tools to address diverse learning needs using multimodal approaches. \\
\hline
Context knowledge -- Communication with Students & Q27 &
I know how to discuss with students the value of generative AI tools while emphasizing critical engagement. &
I know how to discuss with students the value of generative AI tools while emphasizing critical engagement. \\
\hline
Context knowledge -- Communication with Students & Q28 &
I know how to assist students in effectively applying generative AI tools to complex academic tasks like thesis development. &
I know how to assist students in responsibly applying generative AI tools to complex academic tasks like thesis development. \\
\hline
Context knowledge -- Communication with Students & Q29 &
I know how to design and facilitate discussions on professional applications of generative AI tools. &
I know how to design and facilitate discussions on professional applications of generative AI tools. \\
\hline
Context knowledge -- Communication with Students & Q30 &
I know how to engage students in discussions about the ethical implications of generative AI tools. &
I know how to engage students in discussions about the ethical implications of generative AI tools. \\
\hline
Context knowledge -- Communication with Students & Q31 &
I know how to design activities and provide resources to help students stay informed about advancements in generative AI tools. &
I know how to design activities or refer students to reliable resources that help them stay informed about advancements in generative AI tools. \\
\hline
Knowledge of Ethical Implications & Q32 &
I understand how generative AI tools may create or exacerbate inequities among students. &
I understand how generative AI tools may create or exacerbate inequities among students. \\
\hline
Knowledge of Ethical Implications & Q33 &
I understand the implications of using generative AI detection tools on privacy and trust in education. &
I understand the implications of using generative AI detection tools on privacy and trust in education. \\
\hline
Knowledge of Ethical Implications & Q34 &
I understand how much energy is required for tasks performed by generative AI tools. &
I understand how much energy is required for tasks performed by generative AI tools. \\
\hline
Knowledge of Ethical Implications & Q35 &
I understand how generative AI tools may influence critical thinking skills in educational contexts. &
I understand how generative AI tools may positively and/or negatively influence critical thinking skills, creativity, and other cognitive skills. \\
\hline
Knowledge of Ethical Implications & Q36 &
I know how to provide transparency on how generative AI tools are used in my work. &
I know how to provide transparency on how generative AI tools are used in my work. \\
\hline
Knowledge of Ethical Implications & Q37 &
I know how to create strategies to evaluate the ethical use of generative AI tools in my students’ work. &
I reflect on how generative AI impacts the purpose of higher education and the rigor of student learning. \\
\hline
Knowledge of Legal Aspects & Q38 &
I understand the copyright implications of using AI-generated content in academic teaching materials. &
I understand the legal and copyright implications of using and sharing teaching materials created with generative AI tools. \\
\hline
Knowledge of Legal Aspects & Q39 &
I know how to address legal concerns when publishing or sharing materials created with generative AI tools. &
Merged with Q38 \\
\hline
Ability to continuously learn & Q40 &
I know how to stay informed about advancements in generative AI tools. &
I know how to stay informed about advancements in generative AI tools. \\
\hline
Ability to continuously learn & Q41 &
I know how to adapt my teaching practices as generative AI tools evolve. &
I know how to adapt my teaching practices as generative AI tools evolve. \\
\hline
\end{tabular}
\caption{Pre and Post Survey Items for AI Literacy (Policy, Intellectual Growth, Communication, Legal, Lifelong Learning).}
\label{tab:prepost2}
\end{table}

\clearpage

\begin{table}[ht]
\centering
\small
\begin{tabular}{|l|c|c|c|}
\hline
\textbf{Scale} & \textbf{Pre $\alpha$} & \textbf{Post $\alpha$} & \textbf{Interpretation} \\
\hline
Knowledge of capacity \& limitations & 0.10 & 0.67 & Unacceptable $\rightarrow$ Questionable \\
Skill using generative AI tools     & 0.90 & 0.94 & Excellent \\
Detect AI-generated content         & 0.77 & 0.86 & Acceptable $\rightarrow$ Good \\
Assess AI output                    & 0.69 & 0.73 & Questionable $\rightarrow$ Acceptable \\
Prompt engineering skills           & 0.86 & 0.80 & Good \\
Context knowledge -- Policy \& Academic Integrity & 0.80 & 0.71 & Good $\rightarrow$ Acceptable \\
Context knowledge -- Ethics \& Intellectual Growth & 0.88 & 0.86 & Good \\
Context knowledge -- Applications in T\&L & 0.81 & 0.84 & Good \\
Context knowledge -- Communication with Students & 0.81 & 0.90 & Good $\rightarrow$ Excellent \\
Ethical implications                & 0.76 & 0.89 & Acceptable $\rightarrow$ Good \\
Legal knowledge                     & 0.83 & NA   & Good $\rightarrow$ NA \\
Continual learning                  & 0.56 & 0.90 & Poor $\rightarrow$ Excellent \\
\hline
\end{tabular}
\caption{Cronbach’s $\alpha$ values for Gen AI literacy scales and subscales (Pre vs.\ Post). 
While most dimensions demonstrated acceptable to excellent internal consistency, further validation with larger samples is needed, 
particularly for enriching the \textit{Legal knowledge} dimension and rethinking how to measure 
\textit{Knowledge of capacity \& limitations} under conditions of conscious incompetence.}

\label{tab:alpha_results_full}
\end{table}

\begin{figure}[h]
    \centering
    \includegraphics[width=1\textwidth]{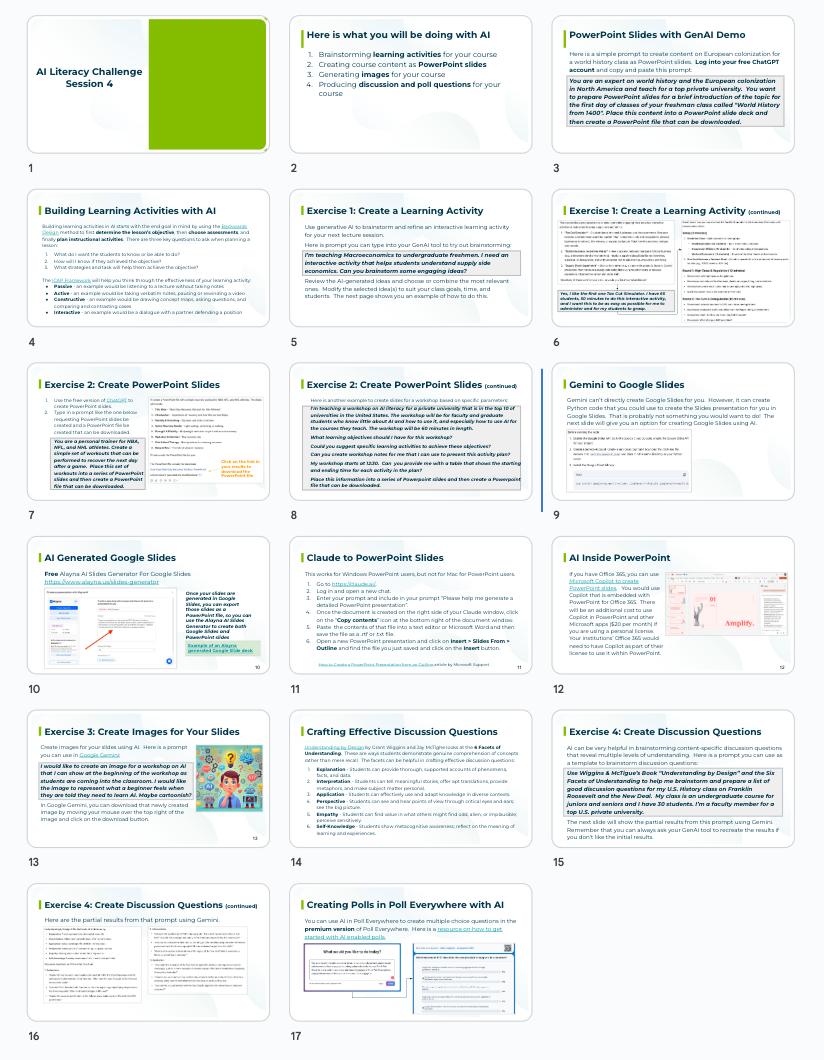} 
    \caption{An Example of the Practice of Using AI Tools}
    \label{fig:session1}
\end{figure}